\documentclass[
aps,prd,
12pt,
nopreprintnumbers,
showpacs,
eqsecnum,
nofootinbib
]{revtex4-1}

\usepackage{amsmath}
\usepackage{amssymb}
\usepackage{graphicx}
\usepackage{bm}

\begin{document}

\preprint{}

\title{
Classical and Quantum Cosmology of Multigravity%
}

\author{Teruki Hanada}\email[]{Deceased}
\author{Koichiro Kobayashi}\email[]{m004wa@yamaguchi-u.ac.jp}
\author{Kazuhiko Shinoda}\email[]{Left physics}
\author{Kiyoshi Shiraishi}\email[]{shiraish@sci.yamaguchi-u.ac.jp}
\affiliation{
Graduate School of Science and Engineering, Yamaguchi University
\\ Yoshida, 
Yamaguchi-shi
, Yamaguchi 753-8512, Japan
}
\date{\today}

\begin{abstract}
Recently, a multigraviton theory on a simple closed circuit graph corresponding to
the discretization of $S^1$ compactification of the Kaluza-Klein (KK) theory has been
considered. In the present paper, we extend this theory to that on a general graph and
study what modes of particles are included. Furthermore, we generalize it in a possible
nonlinear theory based on the vierbein formalism and study classical and quantum
cosmological solutions in the theory.
We found that scale
factors in a solution for this theory repeat acceleration and 
deceleration. 
\end{abstract}

\maketitle


\section{Introduction}
Both astronomical and cosmological data seem to require the presence of yet directly
undetected dark matter and dark energy in the universe. The necessity for these
mysterious components occurs at distances where the gravitational interaction is not
understood sufficiently. This suspicious coincidence inspires a search for
modifications of the  general relativity at large distances.
It is important 
to study massive and multigraviton theory
for understanding cosmology and unification.
In the linear-field theory, gravitons have the Fierz-Pauli (FP) type
masses~\cite{1FP}. But there is an ambiguity in its nonlinear generalization.
We studied thus far the linear multigraviton theory on a circle corresponding to $S^1$
compactification of the KK theory with dimensional deconstruction~\cite{2KS}. 
This model is an extended version of Hamamoto's model~\cite{3Hamamoto} for a massive
graviton.

In this paper,  we construct the FP Lagrangian for multigravitons associated with a
general graph and investigate what modes of particles are included. Furthermore, we
extend it to nonlinear theory based on the vierbein formalism~\cite{N1,N2}.
Nonlinear extensions of multigraviton theory have been studied many authors~\cite{MG}.
In the present paper we focus on the semiclassical sector of the theory which governs
the evolution of the universe; in other words, we will not consider nonlocal
contributions and terms with higher derivatives in the possible complete theory here.

The features of our
model are following: (i) Gravitons as the fluctuation from Minkowski
space-time have the FP type masses~\cite{1FP}. (ii) This model is based on a
generalized dimensional deconstruction method. So, the mass spectrum in the model 
can be tuned more
easily than in the KK theory. (iii) The mass term has a reflection symmetry assigned at each
vertex and an exchange symmetry assigned at each edge of a graph.

In this paper, beginning with graph theoretical description, we introduce
the dimensional deconstruction
\cite{ACG,HPW} and description of the linear theory of multigravity as the basis of our
model in Sec.~\ref{sec2}.  A nonlinear extension of the model is proposed in
Sec.~\ref{sec3}. In Sec.~\ref{sec4}, we consider the vacuum cosmological solutions of
the case associated with the four-site star graph and the four-site path graph. The
study on the quantum cosmological model is exhibited in Sec.~\ref{sec5}. Finally, we
summarize our work and give remarks about the outlook in Sec.~\ref{sec6}.

\section{Multigraviton theory on a general graph}
\label{sec2}

\subsection{FP on a graph}
We consider the matrix representation of the graph theory.%
\footnote{Please see
\cite{refjmp} for a brief review of application of graph theory to field theory, and
textbooks~\cite{GR,CRS} for algebraic graph theory.} A graph $G$ is a pair of $V$ and
$E$, where
$V$ is a set of vertices (sites) while
$E$ is a set of edges (links). 
An edge connects two vertices; two vertices located at the
ends of an edge $e$ are denoted as $o(e)$ and $t(e)$.
Then, we introduce two matrices, an incidence matrix and a graph
Laplacian, associated with a specific graph. 
The incidence matrix $E$ represents the
condition of connection or structure of a graph, and the graph Laplacian $\Delta$ can
be obtained by  $E E^T$, where $E^T$ is the transposed matrix of $E$. By use of these
matrices,  a quadratic form of vectors
$a^T
\Delta a(=a^T E E^T a)$ can be written as a sum of $(a_{t(e)}-a_{o(e)})^2$.  If all
$a_i~(i=1, 2,
\dots, \#V)$, the components of $a$, take the same value, $E^T a=0$ and then
$\Delta a=0$.

So, we consider the Lagrangian for massive gravitons $h^v_{\mu\nu}$ on each vertex 
with the St\"uckelberg vector fields $A^e_{\mu}$
on each edge and a scalar field $\phi^v$ on each vertex:
\begin{eqnarray}
L_m&=&
L_0-\frac{m^2}{2}\sum_{v\in V}\left[h^{v\mu\nu}(EE^T h_{\mu\nu})^v
-h^v(EE^Th)^v\right]\nonumber \\
& &\qquad
-2\sum_{v\in V}\left[m(E A_\mu)^v+\partial_\mu \phi^v\right]
(\partial_\nu h^{v\mu\nu}-\partial^\mu h^v)-\frac{1}{2}\sum_{e\in E}
\left(\partial_\mu A^e_\nu -\partial_\nu A^e_\mu\right)^2\,,
\end{eqnarray}
where $L_0$ is the linearized Einstein-Hilbert Lagrangian:
\begin{equation}
L_0=\sum_{v\in V}\left[-\frac{1}{2}\partial_\lambda h^v_{\mu\nu}\partial^\lambda h^{v\mu\nu}
+\partial_\lambda h^{v\lambda}_{\ \ \mu}\partial_\nu h^{v\nu\mu}
-\partial_\mu h^{v\mu\nu}\partial_\nu h^v+\frac{1}{2}\partial_\lambda h^v 
\partial^\lambda h^v\right]\,,
\end{equation}
and $h^v\equiv\eta^{\mu\nu}h^v_{\mu\nu}$.

This action is invariant under the following transformations:
\begin{equation}
h^v_{\mu\nu}\rightarrow h^v_{\mu\nu}+\partial_\mu \xi^v_\nu +\partial_\nu \xi^v_{\mu},
\quad A^e_\mu \rightarrow A^e_\mu +m(E^T \xi_\mu)^e -\partial_\mu \zeta^e,
\quad \phi^v\rightarrow \phi^v + m(E\zeta)^v,
\end{equation}
where $\xi^v$ and $\zeta^e$ are parameters on each vertex and each edge respectively.
The massive modes of vector and scalar fields are absorbed by the massive modes
of graviton fields due to the symmetry \`a la St\"uckelberg.

Now we examine the gauge fixing of the Lagrangian.
Suppose the following gauge fixing terms:
\begin{eqnarray}
L_{gf}&=&-\sum_{v\in V}\Big[\partial_\nu h^{v\mu\nu}-\frac{1}{2}\partial^\mu
h^v-m(EA^\mu)^v -\partial^\mu \phi^v\Big]^2\nonumber \\
& & -\sum_{e\in E}\Big[\partial_\mu
A^{e\mu}-\frac{m}{2}(E^Th)^e +m(E^T\phi)^e\Big]^2\,,
\end{eqnarray}
then, the gauge-fixed Lagrangian becomes 
\begin{eqnarray}
L_m+L_{gf}&=&\frac{1}{2}H^{\mu\nu}(\partial^2-m^2EE^T)
\Big(H_{\mu\nu}-\frac{1}{2}H\eta_{\mu\nu}\Big)\nonumber \\
& &+A^\mu(\partial^2-m^2E^TE)A_\mu+3\phi(\partial^2-m^2EE^T)\phi\,,
\label{gfl}
\end{eqnarray} 
where
$H_{\mu\nu}=h_{\mu\nu}+\phi\eta_{\mu\nu}$. Here the indices $v$ and $e$, and the
notion of sum over them are omitted.

In the next section, we will see that the mass spectra of fields in the Lagrangian 
for specific graphs with large number of vertices
are similar to those of a five-dimensional model with a compactified extra space.

\subsection{Dimensional deconstruction}
It is assumed that we put fields on vertices or edges. An idea that there are  four
dimensional fields on the sites (vertices) and links (edges), dubbed as dimensional
deconstruction, is introduced by Arkani-Hamed {\it et~al}.~\cite{ACG,HPW}. In this
scheme, the square of mass matrix is proportional to the Laplacian of the
associated graph. 

\begin{figure}[h]
\centering
\includegraphics[height=4cm]
{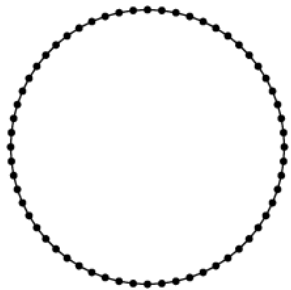}
\caption{%
The cycle graph $C_{60}$.}
\label{C60}
\end{figure}

In the case of a cycle graph (a `closed circuit') with $N$ sites (denoted as $C_N$,
and $C_{60}$ is shown in Fig.~\ref{C60} for example), when
$N$ becomes large, the model on the graph coincides with the five-dimensional theory
with $S^1$ (circle) compactification. In other words, the mass scale of the model
$f$ over
$N$ corresponds to the inverse of the 
circumference 
of the circle:
\begin{equation}M_\ell^2=4f^2(\sin\pi\ell/N)^2\quad \rightarrow\quad  M_\ell^2=(2\pi\ell/L)^2,
\qquad (f/N \rightarrow 1/L)\,.\label{ln}\end{equation}

The mass spectrum is given by the eigenvalues of the graph Laplacian of $C_N$,
which can be expressed as
\begin{equation}
\Delta=\left(
\begin{array}{rrrrr}
2 & -1 &  0& \cdots & -1\\
-1 & 2 & -1 &\cdots &0\\
0 & -1 & 2 &\cdots & 0 \\
\vdots & \vdots & \vdots & \ddots  & \vdots \\
-1 & 0 & 0 & \cdots & 2
\end{array}
\right)\,.
\end{equation}

For a cycle graph, the linear graviton model presented in the previous subsection
coincides with the model proposed in Ref.~\cite{2KS}. The model is
a most general linear multigraviton theory on a generic graph.

\subsection{Particle content in the multigraviton theory on a graph}
\begin{figure}[ht]
\centering
\includegraphics[height=4cm]
{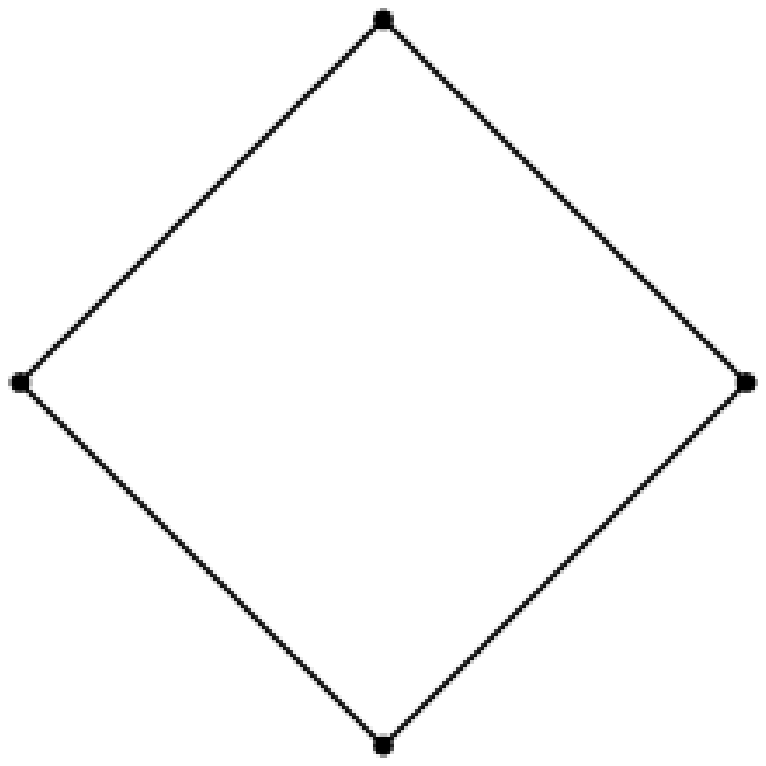}
\qquad\qquad
\includegraphics[height=4cm]
{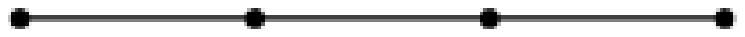}
\caption{%
The cycle graph $C_4$ and the path graph $P_4$.
}
\label{CP}
\end{figure}

For this model, we investigate what modes of particles are contained. Although any
graph is available for the model, here we consider
 two types, a cycle graph $C_N$ and a path graph $P_N$. 
The path graph
has a simple structure like a chain, and has two ends
$(i=1 {\rm ~and~} N)$ and the $i$-th vertex are adjacent to $(i-1)$-th  and $(i+1)$-th
vertices
$(1<i<N)$. 
For example, we show $C_4$ and $P_4$ in Fig.~\ref{CP}. 
The incidence matrix for $C_4$ is defined as
\begin{equation}
E(C_4)=\left(
\begin{array}{rrrr}
1 & 0 &  0 & -1\\
-1 & 1 &0 & 0\\
0 & -1 & 1 & 0 \\
0 & 0 & -1 & 1
\end{array}
\right)\,,
\end{equation}
and then
\begin{equation}
E(C_4)E(C_4)^T=\left(
\begin{array}{rrrr}
2 & -1 &  0 & -1\\
-1 & 2 & -1 & 0\\
0 & -1 & 2 & -1 \\
-1 & 0 & -1 & 2
\end{array}
\right)=E(C_4)^TE(C_4)\,.
\end{equation}
The eigenvalues of $EE^T$ are $\{0, 2, 2, 4\}$ for $C_4$.

On the other hand, 
the incidence matrix for $P_4$ is given by
\begin{equation}
E(P_4)=\left(
\begin{array}{rrr}
1 & 0 &  0 \\
-1 & 1 &0 \\
0 & -1 & 1  \\
0 & 0 & -1 
\end{array}
\right)\,.
\end{equation}
Thus
\begin{equation}
E(P_4)E(P_4)^T=\left(
\begin{array}{rrrr}
1 & -1 &  0 & 0\\
-1 & 2 & -1 & 0\\
0 & -1 & 2 & -1 \\
0 & 0 & -1 & 1
\end{array}
\right)\,,
\end{equation}
and
\begin{equation}
E(P_4)^TE(P_4)=\left(
\begin{array}{rrr}
2 & -1 &  0 \\
-1 & 2 & -1 \\
0 & -1 & 2 
\end{array}
\right)\,,
\end{equation}
are different in their sizes.
The eigenvalues of $EE^T$ are $\{0, 2-\sqrt{2}, 2, 2+\sqrt{2}\}$ and those of $E^TE$
are $\{2-\sqrt{2}, 2, 2+\sqrt{2}\}$ for
$P_4$.
For $P_N$, it is known that the Laplacian eigenvalues are
$4\sin^2\frac{k\pi}{2N}\quad (k=0, 1,\ldots N-1)$.
If we introduce a mass scale $f$ and consider the large $N$ limit as in (\ref{ln}), we
find
$4f^2\sin^2\frac{k\pi}{2N}\rightarrow \left(\frac{\pi k}{L}\right)^2$
where $f/N=1/L$. This spectrum corresponds to that of the compactification on 
$S^1/Z_2$, where the circumference of $S^1$ is $2L$.

In the multigraviton theory associated with the cycle
graph $C_N$  ($\#V=N,$ $\#E=N$),  $N-1$ massive spin-two's, a massless spin-two,
$N-1$ massive vectors, a massless vector, $N-1$ massive scalars, and
a massless scalar seem to be included, as seen from the gauge-fixed
Lagrangian~(\ref{gfl}). The mass spectra of different spin fields are the same, except
for zero modes. This is due to the fact that eigenvalues of $EE^T$ and ones of $E^TE$
are the same except for zero eigenvalues. 

However,   $N-1$ massive spin two, a massless spin two, a
massless vector,  and a massless scalar are left physically, because massive vectors
and massive scalars are absorbed by massive spin two fields to form massive gravitons
with five degrees of freedom each.
 
 Similarly, in the model associated with the path graph $P_N$ ($\#V=N,$
$\#E=N-1$),
$N-1$ massive spin two's, a massless spin two, and a massless scalar is left
physically, the massless vector mode is absent.

The limits of $N$ to infinity in the cases of
$C_N$ and $P_N$ realize the KK theory with $S^1$ and $S^1/Z_2$ compactification,
respectively.

%

\section{Nonlinear extension of a multigraviton theory on a tree graph}
\label{sec3}
Now we will consider a nonlinear extension of the linear theory. Following 
Nibbelink {\it et~al}.~\cite{N1,N2}, we introduce a useful `tool':
\begin{equation}
\langle ABCD \rangle\equiv -\varepsilon_{abcd}\varepsilon^{\mu\nu\rho\sigma}A^a_\mu
B^b_\nu C^c_\rho D^d_\sigma,
\end{equation} 
where $\varepsilon$ is the totally antisymmetric
tensor. Using this expression, we have the Einstein-Hilbert term replacing $A$ and $B$
by vierbeins and $C$ and $D$ by the curvature 2-form. In addition, because the fourth
power of vierbein in the angle  bracket
 is equal to the determinant of vierbeins ($\langle eeee\rangle=\langle
e^4\rangle=24|e|$), this expression means that the Einstein-Hilbert term and the
cosmological term  have the similar structure.

We now assume that the following term is assigned for each edge of a graph:
\begin{equation}\langle (e_1e_1 -e_2e_2)^2 \rangle,\end{equation}
where $e_1$ and $e_2$ are vierbeins at two ends of one edge. Note that this term has a
reflection symmetry $e\leftrightarrow -e$  at each vertex and an exchange symmetry $e_1
\leftrightarrow e_2$ at each edge.

In the weak field limit, {\it i.e.} $e_1=\eta +f_1$, $e_2=\eta+ f_2$,
\begin{equation}\langle(e_1e_1-e_2e_2)^2\rangle=
8\left(\left(\left[f_1\right]-\left[f_2\right]\right)^2
-\left[\left(f_1-f_2\right)^2\right]\right)+O(f^3)\,,\end{equation}
where $\eta$ is the Minkowski metric, and $[f]={\rm tr}f$ for notational simplicity.
This quadratic term corresponds to the FP mass term.\footnote{It is known that the
asymmetric part of $f$ can be omitted~\cite{refbiz}.}

On the other hand, the Einstein-Hilbert term $\frac{1}{2}|e|R$ contains the kinetic
terms of a graviton in the lowest order up to the total derivative: 
\begin{equation}
\frac{1}{2}|e|R
=
-\frac{1}{2}\partial_\lambda f_{\mu\nu}\partial^\lambda f^{\mu\nu}
+\partial_\lambda f^\lambda_{\ \mu}\partial_\nu f^{\nu\mu}
-\partial_\mu f^{\mu\nu}\partial_\nu f
-\frac{1}{2}\partial_\lambda f\partial^\lambda f +O(f^3)\,,\end{equation}
and $\frac{1}{2}R$ contains the following terms in the first order:
\begin{equation}\frac{1}{2}R
=
-\partial^\lambda \partial_\lambda f+\partial_\mu 
\partial_\nu f^{\mu\nu}+O(f^2)\,.\end{equation}

In the case of a tree graph (a graph with no closed circuit---the path graph $P_N$ is
a tree graph, for example), we have the nonlinear Lagrangian of multigraviton theory 
without higher derivative and nonlocal terms,
\begin{equation}L_m=\frac{1}{2}\exp\Phi\sum_{v\in
V}\left|e_v\right|R_v+\frac{M^2}{24}\sum_{e\in E}
\left\langle\left(e_{o(e)}e_{o(e)}-e_{t(e)}e_{t(e)}\right)^2\right\rangle,
\end{equation}
where $R_v$ is the scalar curvature associated with $e_v$ and $M^2\equiv 3 m^2/2$.
The scalar zero-mode field $\Phi$ can be identified as $\phi_1=\phi_2=\cdots
=\Phi$.

\section{Classical cosmology of the multigraviton theory}
\label{sec4}
\begin{figure}[h]
\centering
\includegraphics[height=4cm]
{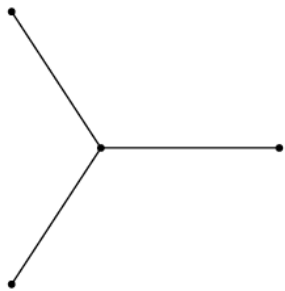}
\caption{%
The star graph $K_{1,3}$.}
\label{K13}
\end{figure}

Now we consider two vacuum cosmological models, associated with a four-site star graph 
and a four-site path graph respectively. Both the star graph and the line graph are
tree graphs. The star graph consists of one central vertex and the other vertices
adjacent to the central one. The star graph $K_{1,3}$ is shown in Fig.~\ref{K13}.
The incidence matrix for $K_{1,3}$ is 
\begin{equation}
E(K_{1,3})=\left(
\begin{array}{rrr}
1 & 1 &  1 \\
-1 & 0 & 0 \\
0 & -1 & 0  \\
0 & 0 & -1 
\end{array}
\right)\,.
\end{equation}
Thus
\begin{equation}
E(K_{1,3})E(K_{1,3})^T=\left(
\begin{array}{rrrr}
3 & -1 &  -1 & -1\\
-1 & 1 & 0 & 0\\
-1 & 0 & 1 & 0 \\
-1 & 0 & 0 & 1
\end{array}
\right)\,,
\end{equation}
and
\begin{equation}
E(K_{1,3})^TE(K_{1,3})=\left(
\begin{array}{rrr}
2 & 1 & 1 \\
1 & 2 & 1 \\
1 & 1 & 2 
\end{array}
\right)\,.
\end{equation}
One can see that the eigenvalues of $EE^T$ are $\{0, 1, 1, 4\}$ and those of $E^TE$ are
$\{1, 1, 4\}$ for the star graph $K_{1,3}$.
For $K_{1,N-1}$, $N$ eigenvalues of the Laplacian are $\{0,1,\ldots,1,N\}$.
The degeneracy of $N-2$ eigenvalues $(=1)$ is apparently due to the symmetry of the
star graph.


In the case of the star graph, the associated Lagrangian for multigravitons is the
following;
\begin{equation}L_{star}=\frac{1}{2}\exp\Phi\sum_{i=1}^4|e_i|R_i+\frac{M^2}{24}\sum_{i=2}^4
\left\langle (e_1e_1-e_ie_i)^2\right\rangle,\end{equation} where, $e_1$ is on the
center of the graph. On the other hand, the Lagrangian of the case of the path graph is
\begin{equation}L_{path}=\frac{1}{2}\exp\Phi\sum_{i=1}^4|e_i|R_i+\frac{M^2}{24}\sum_{i=1}^{3}\left
\langle (e_ie_i-e_{i+1}e_{i+1})^2\right\rangle,\end{equation} where, $e_1$ and $e_4$
are on each end of the graph.

Now let us introduce the setting for cosmology.
We assume the homogeneous universe with a spatially-constant scalar field $\Phi(t)$
and  the following metric;
\begin{equation}g_{\mu\nu}dx^\mu dx^\nu=-e^{-\Phi(t)}dt^2+e^{-\Phi(t)}A_i^2(t)
(dr^2+r^2d\Omega^2)\,,
\end{equation} 
where 
$A_i(t)~(i=1,\cdots, 4)$ are scale factors. Then, 
\begin{equation}\left\langle (e_ie_i-e_je_j)^2\right\rangle=e^{-2\Phi(t)}
(e^{a_i(t)}-e^{a_j(t)})(e^{2a_i(t)}-e^{2a_j(t)})\,,\end{equation}
where $a_i(t)\equiv \ln A_i(t)$.

\begin{figure}[ht]
\centering
\includegraphics[height=3cm]
{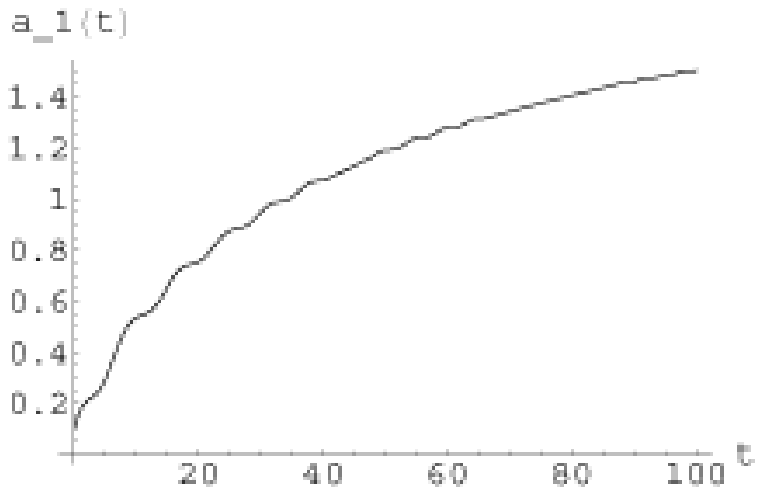}
\qquad\qquad
\includegraphics[height=3cm]
{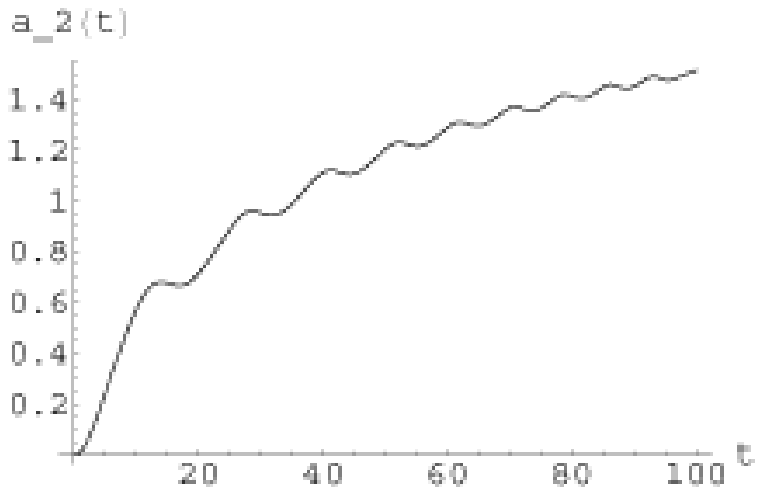}\\
\includegraphics[height=3cm]
{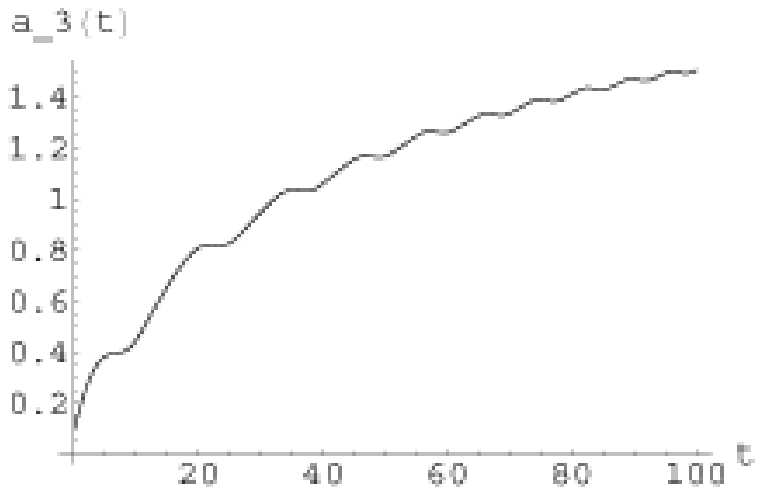}
\qquad\qquad
\includegraphics[height=3cm]
{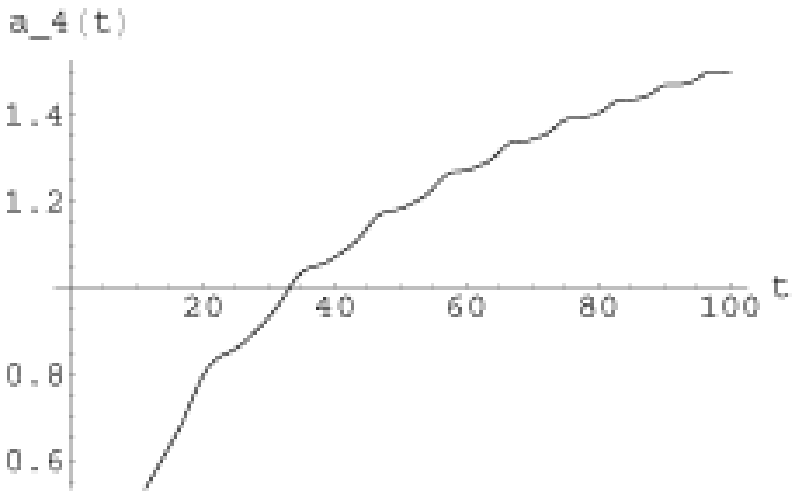}\\
\includegraphics[height=3cm]
{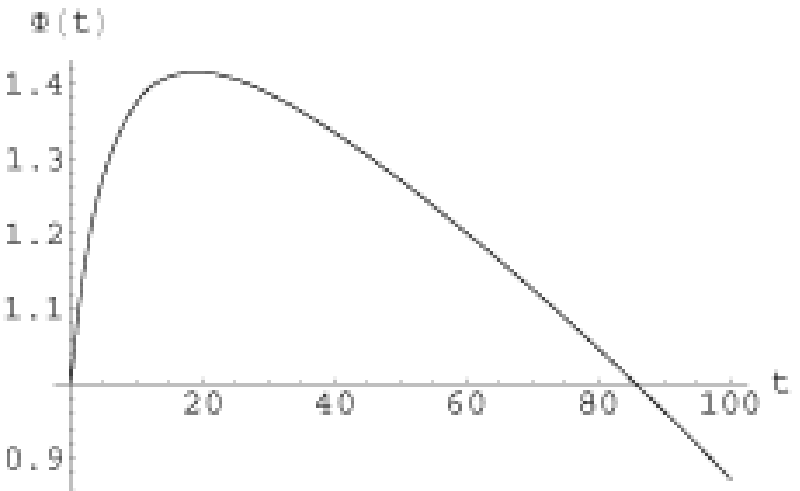}
\caption{%
Numerical solutions of  
$a$'s and $\Phi$
in the case of the four-site star
graph.}
\label{fig1}
\end{figure}

\begin{figure}[ht]
\centering
\includegraphics[height=3cm]
{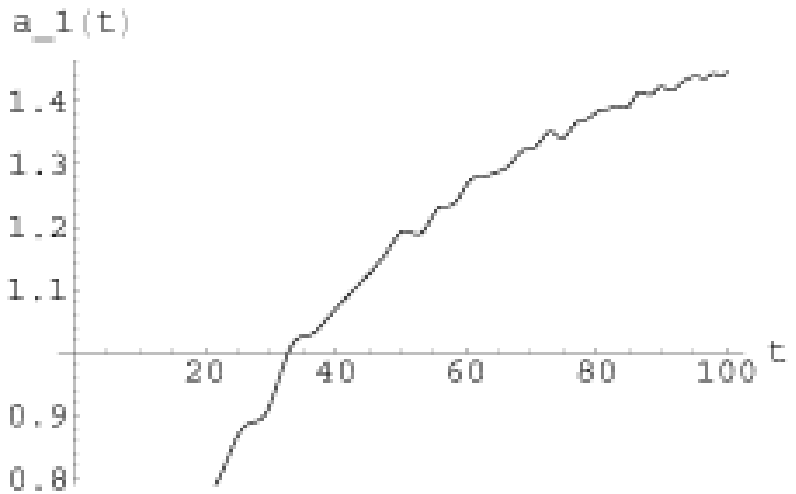}
\qquad\qquad
\includegraphics[height=3cm]
{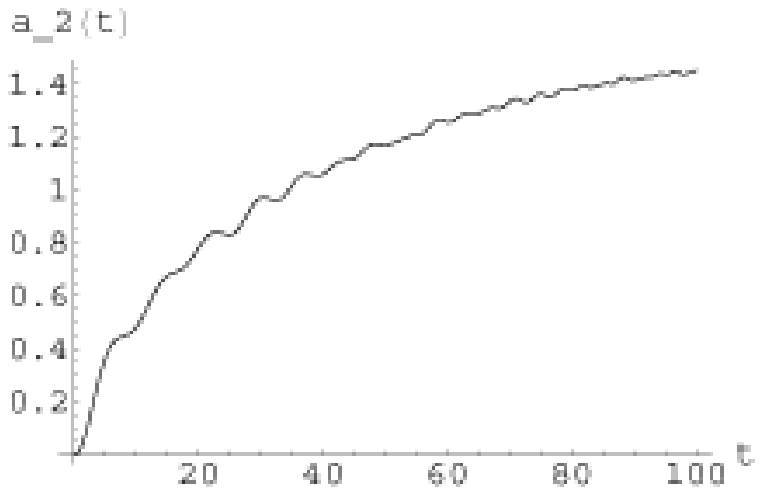}\\
\includegraphics[height=3cm]
{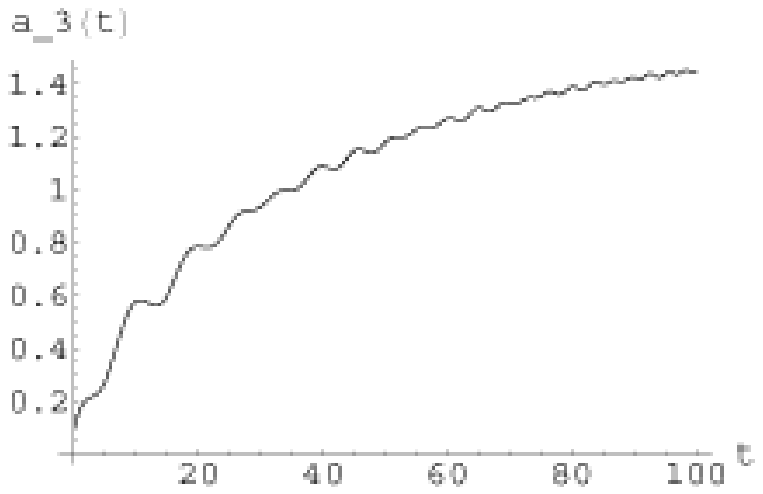}
\qquad\qquad
\includegraphics[height=3cm]
{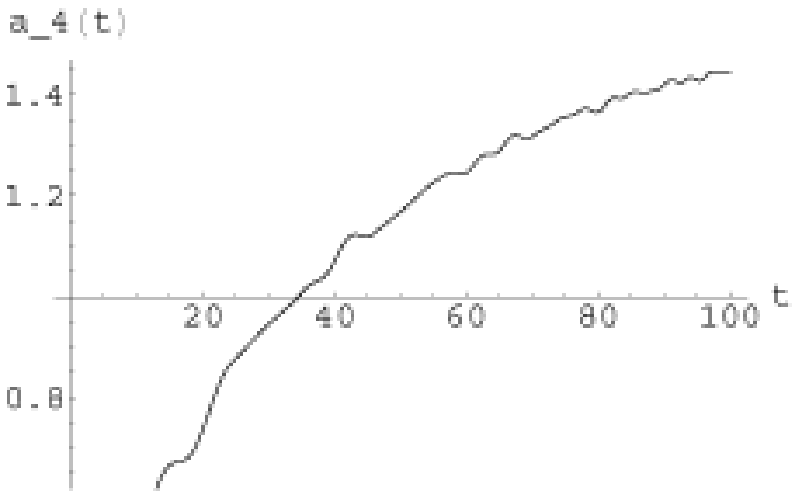}\\
\includegraphics[height=3cm]
{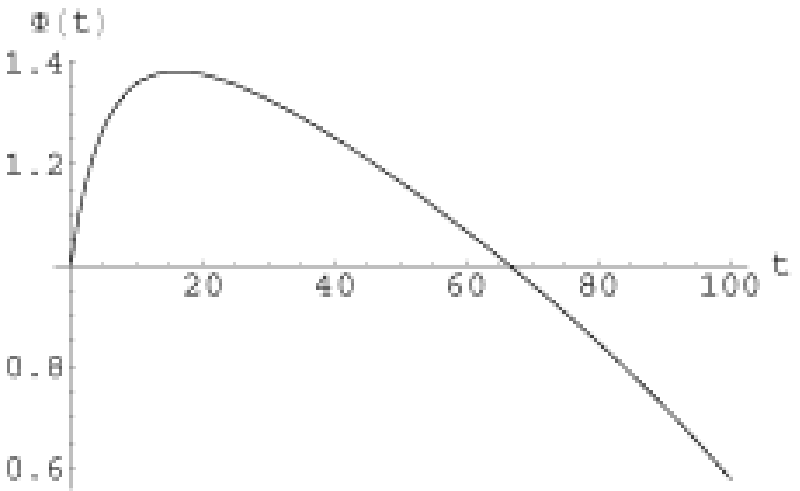}
\caption{%
Numerical solutions of  
$a$'s and $\Phi$
in the case of the four-site path
graph.}
\label{fig2}
\end{figure}

We show the results of numerical calculations for the two models
on the same appropriate initial conditions in Fig.~\ref{fig1} and Fig.~\ref{fig2}.
In both cases the scalar field $\Phi$ behaves similarly and in each case scale factors
repeat the increase and the decrease. 
The oscillation of the scale factors in
the path graph case  include more different modes than that of the scale factors in the
case of the star graph where the degeneracy of eigenvalues exists.

The star graph model has more symmetries than the path graph model. 
Therefore a lot of modes in the star graph are degenerate, while there is no
degeneracy in the spectrum of the line graph.  In the path graph case, increase of
the number of sites gives the more complicated behaviors of the scale factors. On
the other hand, in the star graph case, the symmetries are preserved even if the number
of sites increases.  Therefore, the behaviors of scale factors are much similar to
those in the four-site model, essentially.

%

\section{Quantum cosmology of the multigraviton theory}
\label{sec5}

\subsection{The Wheeler-DeWitt equation}
In the previous section, we have seen the oscillatory behavior in the evolution of
scale factors. As a qualitative analysis, we only show the characteristic solutions.
In fact, oscillations must be dependent on the initial conditions.
What are the natural conditions? To study the initial state, we have to consider
quantum behavior of cosmology.  Note that quantum cosmology of multigraviton theory
has never been studied yet as far as we know.

\begin{figure}[ht]
\centering
\includegraphics[height=3cm]
{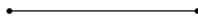}
\caption{%
The path graph $P_{2}$.}
\label{P2}
\end{figure}

In this section we consider a minimal model based on a graph $P_2$,
which is shown in Fig.~\ref{P2}
This model has two
gravitons,%
\footnote{In this case, the eigenvalues of mass are $0$ and $2M/\sqrt{3}$.}
 or two scale factors.
The Lagrangian density is given by
\begin{equation}{\cal
L}=\frac{1}{2\kappa^2}\exp\Phi\left[|e_g|R_g+|e_f|R_f\right]
+\frac{M^2}{24}\langle(e_ge_g-e_fe_f)^2\rangle+{\rm (surface~terms)}\,,
\end{equation}
where two graviton fields are labeled by $g$ and $f$.
This model in this case is very similar to $f$-$g$ gravity~\cite{fg} 
or bigravity~\cite{big},
but our model also contains a massless scalar field.

We take the metric ansatze as follows:
\begin{equation}ds_f^2=e^{-\Phi}(-N^2dt^2+A^2 d\Omega_3^2)\,,
\end{equation}
\begin{equation}ds_g^2=e^{-\Phi}(-N^2dt^2+B^2 d\Omega_3^2)\,.
\end{equation}
These choices are equivalent to field redefinitions
$g_{\mu\nu}=e^{-\Phi}g^{(E)}_{\mu\nu}$ so that
\[{\cal
L}=\frac{1}{2\kappa^2}\left[|e^{(E)}_g|R^{(E)}_g+|e^{(E)}_f|R^{(E)}_f\right]+\cdots
\]
and often quoted as the choice of the Einstein frame.
Here we assume that $A$, $B$ and $\Phi$ depend only on
$t$, and $d\Omega_3^2=dx^2+dy^2+dz^2$. 
The lapse function will be set as $N=1$ after the calculation (by the
redefinition of $t$).   Each metric is homogeneous, isotropic, and flat 
in the Einstein frame, in the present
analysis. 
Then the action reads
\begin{eqnarray}\int L dt&=&\int
dt\left[\frac{3}{2N\kappa^2}
\left\{e^{3\alpha}(-4\dot{\alpha}^2+\dot{\Phi}^2)+e^{3\beta}
(-4\dot{\beta}^2+\dot{\Phi}^2)\right\}\right.\nonumber \\& &\qquad\left.\frac{ }{
}+NM^2e^{-2\Phi}(e^{\alpha}+e^{\beta})(e^{\alpha}-e^{\beta})^2\right]\,,
\end{eqnarray}
where $\alpha=\ln A$\,, and $\beta=\ln B$. The dot indicates the derivative with
respect to $t$. The conjugate variables are
\begin{equation}\pi_\alpha=-\frac{12e^{3\alpha}\dot{\alpha}}{N\kappa^2}\,,
\qquad\pi_\beta=-\frac{12e^{3\beta}\dot{\alpha}}{N\kappa^2}\,,
\qquad\pi_\Phi=\frac{3(e^{3\beta}+e^{3\beta})\dot{\Phi}}{N\kappa^2}\,,
\end{equation}
thus we obtain the Hamiltonian of the universe as
\begin{equation}H=N\left[-\frac{\kappa^2}{24}
\left\{e^{-3\alpha}\pi_\alpha^2+e^{-3\beta}\pi_\beta^2
\right\}+\frac{\kappa^2}{6(e^{3\alpha}+e^{3\beta})}\pi_\Phi^2
-M^2e^{-2\Phi}(e^{\alpha}+e^{\beta})(e^{\alpha}-e^{\beta})^2\right]\,.
\end{equation}

From the Hamiltonian, we obtain the Wheeler-DeWitt (WDW) equation for the wave
function of the universe $\Psi(\alpha,\beta,\Phi)$~\cite{Halliwell}.
Although there are ambiguities in the ordering, we adopt the simple replacement of
conjugate variables by the derivatives with respect to the corresponding dynamical
variables.\footnote{Another plausible choice is adoption of the Laplacian in the
minisuperspace. The qualitative behavior is not changed by the choice of the
operator orderings.}
The WDW equation for the present model is
\begin{eqnarray}& &\left[-\frac{\kappa^2}{24}
\left\{e^{-3\alpha}\frac{\partial^2}{\partial\alpha^2}+e^{-3\beta}
\frac{\partial^2}{\partial\beta^2}
\right\}+\frac{\kappa^2}{6(e^{3\alpha}+e^{3\beta})}
\frac{\partial^2}{\partial\Phi^2}\right.\nonumber \\& &\left.\frac{}{}+M^2e^{-2\Phi}
(e^{\alpha}+e^{\beta})(e^{\alpha}-e^{\beta})^2\right]\Psi(\alpha,\beta,\Phi)=0\,.
\label{WDWO}
\end{eqnarray}

Now we introduce new variables $x$ and $y$. They are defined as
\begin{equation}x=\frac{\alpha+\beta}{2}\,,\qquad y=\frac{\alpha-\beta}{2}\,.
\end{equation}
Since
\begin{equation}\frac{\partial}{\partial\alpha}=\frac{1}{2}\left(\frac{\partial}{\partial
x}+\frac{\partial}{\partial
y}\right)\,,\qquad
\frac{\partial}{\partial\beta}=\frac{1}{2}\left(\frac{\partial}{\partial
x}-\frac{\partial}{\partial y}\right)\,,
\end{equation}
the WDW equation (\ref{WDWO}) is rewritten as
\begin{eqnarray}& &\left[-\frac{\kappa^2}{6}
\left\{\cosh 3y\, (\frac{\partial^2}{\partial
x^2}+\frac{\partial^2}{\partial y^2})-2\sinh 3y
\frac{\partial^2}{\partial x\partial y}
\right\}+\frac{\kappa^2}{6\cosh 3y}
\frac{\partial^2}{\partial\phi^2}\right.\nonumber \\& &\left.\frac{}{}
+64M^2e^{-4\phi}e^{6x}\cosh y \sinh^2 y\right]\Psi(x,y,\phi)=0\,,
\label{WDW}
\end{eqnarray}
where we have also introduced $\phi=\Phi/2$ for simplicity.

\subsection{Wave-packet solutions}

To analyze the WDW equation (\ref{WDW}), we assume the wave packet ansatz.
The wave packet in quantum cosmology was originally introduced in the
references~\cite{Kazama,Kie}, and is utilized recently for various models
such as in Ref.~\cite{Kamenshchik}.
The use of the wave packet is crucial for the case with no special `initial' state
served as in the case with the positively curved homogeneous space.

The general form of the wave function is written by
\begin{equation}\Psi(x,y,\phi)=\sum_n C_n(x,\phi)\psi_n(x,y,\phi)\,,
\end{equation}
where
\begin{equation}\left[-\frac{\kappa^2}{6}
\frac{\partial^2}{\partial y^2}+64M^2e^{-4\phi}e^{6x}\frac{\cosh
y
\sinh^2 y}{\cosh 3y}\right]\psi_n(x,y,\phi)=E_n(x,\phi)\psi_n(x,y,\phi)\,.
\end{equation}
We assume that $x$ and $\phi$ are slowly evolving variables, while $y$ is a rapidly
changing variable.%
\footnote{This assumption leads to a universe with the increasing mean size, which
looks like our present universe. Some violent evolutions can occur in the very
early universe, but we do not consider the possiblity here.}
In other words, we assume
$\partial_x\ln C_n\ll 1$ and $\partial_\phi\ln C_n\ll 1$.

Further we approximate the equation if $y$ has a small amplitude. Then
\begin{equation}\left[-\frac{\kappa^2}{6}
\frac{\partial^2}{\partial y^2}
+64M^2e^{6x-4\phi}y^2\right]\psi_n(x,y,\phi)=E_n(x,\phi)\psi_n(x,y,\phi)\,.
\label{eq1}
\end{equation}
If $x$ and $\phi$ are slowly-developing variables, this is no other than
the equation for a harmonic oscillator. 
The differential equation
\begin{equation}\psi''(y)-by^2\psi(y)+c\psi(y)=0\,
\end{equation}
has the solution
\begin{equation}\psi_n(y)=\exp\left(-\frac{1}{2}\left[\sqrt{b}\right]y^2\right){\rm
H}_n({ \left[b\right]^{1/4}}y)\,,
\end{equation}
where ${\rm H}_n$ is Hermite polynomial in the definition of {\it Mathematica} and
$\psi$ is normalizable if
\begin{equation}c=c_n=2\sqrt{b}\left(n+\frac{1}{2}\right)\,,\qquad
n={\rm integer}
\label{eq2}
\end{equation}

Therefore the approximation gives the solution of (\ref{eq1}) which leads to
\begin{equation}E_n(x,\phi)=\frac{\kappa^2}{6}c_n\,,\end{equation}
where $c_n$ is given by (\ref{eq2}) with
\begin{equation}
b=\frac{6}{\kappa^2}64M^2 e^{6x-4\phi}\,.
\end{equation}

Now the differential equation for $C_n$ becomes
\begin{equation}\left[-\frac{\kappa^2}{6}
\frac{\partial^2}{\partial
x^2}
+\frac{\kappa^2}{6}
\frac{\partial^2}{\partial\phi^2}+E_n(x,\phi)\right]C_n(x,\phi)=0\,,
\end{equation}
and can be approximated as 
\begin{equation}\left[
\frac{\partial^2}{\partial
x^2}
-
\frac{\partial^2}{\partial\phi^2}-16M\sqrt{\frac{6}{\kappa^2}
}\left(n+\frac{1}{2}\right)e^{3x-2\phi}\right]C_n(x,\phi)=0\,.
\end{equation}

Further rewriting variables as
\begin{equation}X\equiv\frac{3x-2\phi}{\sqrt{5}}\,,\qquad
Z\equiv\frac{3\phi-2x}{\sqrt{5}}\,,
\end{equation}
leads to
\begin{equation}\left[
\frac{\partial^2}{\partial
X^2}
-
\frac{\partial^2}{\partial Z^2}-16M\sqrt{\frac{6}{\kappa^2}
}\left(n+\frac{1}{2}\right)e^{\sqrt{5}X}\right]{C}_n(X,Z)=0\,.
\label{eq3}
\end{equation}

Finally, separating variables as
${C}_n(X,Z)=f_k(Z)\varphi_{kn}(X)$ according to
\begin{equation}\left[
\frac{\partial^2}{\partial Z^2}+k^2\right]f_k(Z)=0\,,
\end{equation}
tells us the solution
\begin{equation}f_k(Z)=e^{-ikZ}\,,\qquad
\varphi_{kn}(X)=K_{2ik/\sqrt{5}}\left[\frac{2\sqrt{a_n}
e^{\sqrt{5}X/2}}{\sqrt{5}}\right]\,,
\end{equation}
where $K_\nu$ is the modified Bessel function of the second kind with
\begin{equation}a_n=16M\sqrt{\frac{6}{\kappa^2}
}\left(n+\frac{1}{2}\right)\,.
\end{equation}

The wave packet can be written in the form
\begin{equation}\Psi=\sum_{n=0}^\infty \psi_n(y)\int_{-\infty}^\infty dk\,
A_n(k) K_{2ik/\sqrt{5}}\left[\frac{2\sqrt{a_n}
e^{\sqrt{5}X/2}}{\sqrt{5}}\right]e^{-ikZ}\,.
\end{equation}

The wave function behaves oscillatory in the region $X<0$ and exponentially damps in
the region $X>0$. This is because the exponential potential `wall' in (\ref{eq3}).
The amplitude with respect to $X$ has a maximum peak at $X\sim 0$ independently to $k$.
Therefore the general wave packet, in which $A(k)$ is taken to be a Gaussian, has a
peak at
$X\sim 0$, because other peaks are destructively superposed.

The universe with $X\sim 0$ is preferred in general.
Even in classical solution,
oscillatory $y$ leads to $x\sim 2/3 \phi$ can be confirmed.

\subsection{Comparison to the case with no oscillation}

If we assume `classically' $y\sim 0$, {\it i.e.}, assume $\alpha=\beta$,
WDW equation  reads
\begin{equation}\left[\frac{\partial^2}{\partial x^2}-\frac{\partial^2}{\partial
\phi^2}\right]\psi=0\,,
\end{equation}
or
\begin{equation}\left[\frac{\partial^2}{\partial X^2}-\frac{\partial^2}{\partial
Z^2}\right]\psi=0\,.
\end{equation}
The solution of this differential equation is:
\begin{equation}\psi=f_1(x-\phi)+f_2(x+\phi)=g_1(X-Z)+g_2(X+Z)\,.
\end{equation}
This shows much different behaviors from the `correct' solution of the WDW equation.
No typical peak can be expected. This is rather trivial, but this comparison reminds
us the fact that there is at least zero-point oscillation in any
oscillatory quantum system.

\section{Conclusion and outlook}
\label{sec6}
We have studied the simple and Lorentz-invariant theory of multigraviton, and have
shown typical cosmological solutions. We focused our attention on the models
associated with the four-site star graph and the  path graph and found that vacuum
cosmological solutions with the scale factors  show the repeated accelerating and
decelerating expansions. The differences between these two models were discussed from
a viewpoint about  symmetries.
By using a simplest model, we also qualitatively showed that the oscillatory behavior
is considered as necessary in quantum universe. We should investigate more plausible
and applicable solutions for classical as well as quantum cosmology, including usual
matter.

To this end, we should study how the gravitons and the scalar field couples to various
matter fields.
%
To consider various coupled fields,
incorporation of supersymmetry or supergravity is also of much interest.
Permitting higher derivative terms and nonlocal terms in the action will
bring more possibilities to the completion of nonlinearity and be worth studying
still.

As the future works, from the mathematical point of view, it is interesting to
construct models with the use of generic graphs, such as weighted graphs, fractals,
and so on.
%

\acknowledgements
We would like to thank N.~Kan for useful comments.
We also would like to thank the organizers of JGRG17 and JGRG18, where our
partial results ({\tt [arXiv:0801.2641]} and {\tt [arXiv:0902.0103]}) were presented.

\bibliographystyle{apsrev4-1}

\begin{thebibliography}{99}
\bibitem{1FP} M.~Fierz and W.~Pauli, Proc. Roy. Soc. Lond. {\bf A173} (1939) 211.
\bibitem{2KS} N.~Kan and K.~Shiraishi, Class. Quant. Grav. {\bf 20} (2003) 4965
[arXiv:gr-qc/0212113].
\bibitem{3Hamamoto} S.~Hamamoto, Prog. Theor. Phys. {\bf 97} (1997) 327
[arXiv:hep-th/9611141].
\bibitem{N1} S.~G.~Nibbelink, M.~Peloso and M.~Sexton, Eur. Phys. J. {\bf C51} (2007)
741 [arXiv:hep-th/0610196].
\bibitem{N2} S.~G.~Nibbelink and M.~Peloso, Class. Quant. Grav. {\bf 22} (2005) 1313
[arXiv:hep-th/0411184].
\bibitem{MG}   
N.~Arkani-Hamed, H.~Georgi and M.~D.~Schwartz,
Ann. Phys. {\bf 305} (2003) 96 [arXiv:hep-th/0210184];
N.~Arkani-Hamed and M.~D.~Schwartz,
Phys. Rev. {\bf D69} (2004) 104001 [arXiv:hep-th/0302110];
M.~D.~Schwartz,
Phys. Rev. {\bf D68} (2003) 024029 [arXiv:hep-th/0303114];
G.~Cognola, E.~Elizalde, S.~Nojiri, S.~D.~Odintsov and S.~Zerbini,
Mod. Phys. Lett. {\bf A19} (2004) 1435 [arXiv:hep-th/0312269];
S.~Nojiri and  S.~D.~Odintsov,
Phys. Lett. {\bf B590} (2004) 295 [arXiv:hep-th/0403162];
F.~Bauer, T.~Hallgren and G.~Seidl,
Nucl. Phys. {\bf B781} (2007) 32 [arXiv:hep-th/0608176];
G.~Seidl, e-Print: arXiv:0901.4304 [hep-th].
\bibitem{ACG} N. Arkani-Hamed, A. G. Cohen and H. Georgi, Phys. Rev. Lett. {\bf 86}
(2001) 4757 [arXiv:hep-th/0104005].
\bibitem{HPW} C.~T.~Hill, S. Pokorski and J. Wang, 
Phys. Rev. {\bf D64} (2001) 1050050 [arXiv:hep-th/0104035].
\bibitem{refjmp} N.~Kan and K.~Shiraishi, 
J. Math. Phys. {\bf 46} (2005) 112301 [arXiv:hep-th/0409268].
\bibitem{GR} C.~Godsil and G.~Royle, {\it Algebraic Graph Theory} (Springer, New York,
2001).
\bibitem{CRS} D.~Cvetkovi\'c, P.~Rowlinson and S.~Simi\'c, 
{\it An Introduction to the Theory of Graph Spectra} (London Mathematical Society
Student Texts 75)  (Cambridge University Press, Cambridge, UK, 2010).
\bibitem{refbiz} C.~Bizdadea et al., JHEP {\bf 02} (2005) 016;
C.~Bizdadea et al., Eur. Phys. J. {\bf C48} (2006) 265.
\bibitem{Halliwell} For a concise review, J.~J.~Halliwell, {``Introductory Lectures on
Quantum Cosmology''}, in {\it Proceedings of the Jerusalem Winter School
on Quantum Cosmology and Baby Universe} (edited by T.~Piran, World Scientific,
Singapore, 1991), 
arXiv:0909.2566[gr-qc].
\bibitem{Kazama} Y.~Kazama and R.~Nakayama, Phys. Rev. {\bf D32} (1985) 2500.
\bibitem{Kie} C.~Kiefer, Phys. Rev. {\bf D38} (1988) 1761.
\bibitem{Kamenshchik} A.~Y.~Kamenshchik, C.~Kiefer and B.~Sandh\"ofer,
Phys. Rev. {\bf D76} (2007) 064032.

\bibitem{fg} 
C.~J.~Isham, A.~Salam and J.~Strathdee, Phys. Rev. {\bf D3} (1971) 867;
A.~Salam and J.~Strathdee, Phys. Rev. {\bf D16} (1977) 2668;
A.~Salam and J.~Strathdee, Phys. Lett. {\bf B67} (1977) 429;
C.~J.~Isham and D.~Storey, Phys. Rev. {\bf D18} (1978) 1047.

\bibitem{big} 
T.~Damour and I.~I.~Kogan, Phys. Rev. {\bf D66} (2002) 104024;
D.~Blas, C.~Deffayet and J.~Garriga, Class. Quant. Grav. {\bf 23} (2006) 1697;
D.~Blas, C.~Deffayet and J.~Garriga, Phys. Rev. {\bf D76} (2007) 104036;
D.~Blas, Int. J. Theor. Phys. {\bf 46} (2007) 2258;
Z.~Berezhiani, D.~Comelli, F.~Nesti and L.~Pilo, Phys. Rev. Lett. {\bf 99} (2007)
131101.


\end{thebibliography}

\end{document}